\documentclass[informatics,article,accept,moreauthors,pdftex,10pt,a4paper]{mdpi} 
\usepackage{color}
\usepackage{cases}
\usepackage{multirow}
\usepackage{graphicx,url}
\usepackage[linesnumbered,ruled,vlined]{algorithm2e}
\usepackage{amsthm}
\usepackage{amsmath}
\usepackage{amssymb}
\usepackage{cancel}

\usepackage{array}

\theoremstyle{definition}
\newtheorem{Def}{Rule}
\newtheorem{defi}{Definition}

\newcommand{\exTrace}{G}
\newcommand{\intBegin}[1]{{#1}_{b}}
\newcommand{\intEnd}[1]{{#1}_{e}}
\newcommand{\HT}[1]{\textcolor{black}{#1}}

\firstpage{1} 
\articlenumber{14}
\doinum{10.3390/informatics5010014}
\pubvolume{5}
\pubyear{2018}
\copyrightyear{2018}
\history{Received: 5 December 2017; Accepted: 2 March 2018; Published: 8 March 2018}
\Title{Utilizing Provenance in Reusable Research Objects}

\Author{Zhihao Yuan 
$^{1}$, Dai Hai Ton That $^{1}$, Siddhant Kothari $^{2}$, Gabriel Fils $^{1}$ and Tanu Malik $^{1, *}$}

\AuthorNames{Firstname Lastname, Firstname Lastname and Firstname Lastname}

\address{%
	$^{1}$ \quad School of Computing, DePaul University, Chicago, IL, 60604,  
	USA;  zhihao.yuan@depaul.edu (Z.Y.); dtonthat@depaul.edu (D.H.T.T.); gfils1@depaul.edu (G.F.);  tmalik1@depaul.edu (T.M.)\\
	$^{2}$ \quad Department of Computer Science, University of Chicago, Chicago, IL, 60637, USA; 	
siddhant22@uchicago.edu (S.H.)
	 }

\corres{Correspondence: tmalik1@depaul.edu; Tel.: +1-362-312-1121} 

\abstract{Science is conducted collaboratively, often requiring the sharing of knowledge about computational experiments. When experiments include only datasets, they can be shared using Uniform Resource Identifiers (URIs) or Digital Object Identifiers (DOIs). An experiment, however, seldom includes only datasets, but more often includes software, its past execution, provenance, and associated documentation. The Research Object has recently emerged as a comprehensive and systematic method for aggregation and identification of diverse elements of computational experiments. While a necessary method, mere aggregation is not sufficient for the sharing of computational experiments. Other users must be able to easily recompute on these shared research objects. Computational provenance is often the key to enable such reuse. {In this paper, we show how reusable research objects can utilize provenance to correctly repeat a previous reference execution, to construct a subset of a research object for partial reuse, and to reuse existing contents of a research object for modified reuse. We describe two methods to summarize provenance that aid in understanding the contents and past executions of a research object. The first method obtains a process-view by collapsing low-level system information, and the second method obtains a summary graph by grouping related nodes and edges with the goal to obtain a graph view similar to application workflow. Through detailed experiments, we show the efficacy and efficiency of our algorithms.}}


\begin{document}
	
	\section{Introduction}
	
	Research objects---aggregations of  digital artifacts such as code, data, scripts, and temporary experiment results---provide a means to share knowledge about computational experiments {~\cite{Miksa:2017:RO, Belhajjame:RO:2015}}. \textls[-10]{In~recent times, sharing computational experiments has become vital; scientific claims, inevitably asserted via computational experiments, remain poorly verified in text-based research papers. Research objects, together with the paper, provide an authoritative and far more complete record of a piece of~research. }
	
	Several tools now exist to help authors create research objects from a variety of digital artifacts (see~\cite{Stodden2014} for several tools and~\cite{Malik:2014:SOLE} for a variety of research objects). The tools enable research objects to be shared on websites that disseminate scholarly information, such as Figshare~\cite{Figshare}. Despite~their advantages, shared research objects do not permit easy reuse of their contents to verify their computations, or easy adaptation of their contents for reuse in new experiments. Often the extent of reuse is subject to the amount of accompanying documentation, which may be limited to compilation and installation instructions. If documentation is scant, research objects will remain unused.
	
	\pagebreak
		
	The~minimum use-case for sharing a computational experiment (in the form of a shared research object) involves repeating its original execution and verifying its results. To truly exploit its potential, however, it must support modified reuse. Therefore, the research object must be created and stored not as a simple aggregation of digital content, as previously advocated~\cite{Belhajjame:RO:2015, RO-Bundle}, but in a readily-computable form: as a {\emph{reusable}}  
	research object. We demonstrate the distinction in two ways.
	
	Consider a typical research paper with an analysis based on large amounts of code and data, and assume that the researcher authoring the paper has used the code and data to conduct a number of experiments that produce the paper's target figures and results. The example paper's digital artifacts relating to its experiments may be bundled together in a medium such as a file archive (.tar), compressed file format (.gz), virtual image, or container. A shared research object is free to use any of these mediums. A reusable research object, however, must use a virtual image or container, since it must produce a computational research object that, when downloaded and shared, will guarantee an instantly-executable unit of computation.
	
	Also consider the example paper's metadata, which, similar to the metadata in most papers, is interspersed throughout the project's written analysis, and throughout its code and data. The~metadata can take many forms, including annotations, version information, and provenance. A shared research object's metadata usually serves a purely informational purpose and is seldom used literally in the paper's experiments.  A reusable research object, however, utilizes literal metadata by directly linking it to the code and data of the experiments. In particular, computational provenance, if collected in standard form, can guide different forms of reusable analysis---exact, partial, or modified reuse. 
	In~other words, a reusable research object can execute conditionally based on its embedded metadata, instead of simply including it as a stand-alone digital artifact that requires more interpretive labor to reason about and reuse. 
	
	{Several tools to create reusable research objects have been recently proposed~\cite{Chirigati:2013:ReproZip,Janin:CARE,Ton:2017:Sciunit}. All tools use application virtualization (AV) to automatically create a container of an executable application. In~AV, operating system calls during application execution are interrupted to enable the copying of all binaries, data, external user input, and software dependencies into a container. The resulting container is portable and instantly reusable: it can be run on any compatible machine without installation, configuration, or root permissions. However, the tools differ in the method of re-execution. In~particular, none except Sciunit~\cite{Ton:2017:Sciunit} captures provenance both during creation of the container and its re-execution (see Section~\ref{sec:research object evolution} for further differences). Capturing provenance during container creation and then at each step of re-execution can be useful for a variety of purposes. In this paper, we show how provenance audited at execution and re-execution time within containers can be utilized to establish exact repetition of a previous reference execution, to construct a subset for reusing part of a research object, and to reuse contents when research objects are re-executed with different inputs.} 
	
	
	Reusable research objects, owing to application virtualization, store an {\emph{execution trace}}.  
To generate provenance  from an execution trace, dependency information must be inferred from the trace. We~show how this information can be inferred in a lazy or post hoc manner. To use inferred provenance within a reusable research object, an important consideration is the granularity at which the execution trace is audited. Auditing 
at the granularity of read and write of each data variables can help detect concurrent processing within application programs but imposes significant run-time overhead. Thus, we audit at the granularity of open and close of files and spawn of processes. At this granularity, given concurrent programs, exact repeatability cannot be guaranteed. For instance, if two processes read and write to the same file at the same time, in the absence of read/write provenance dependence information, the order in which they wrote to the file cannot be guaranteed. To still use provenance for container re-execution, we do not break cycles due to concurrent processing, but assume that applications are willing to re-run a few extra processes.  
	
	
	Even if provenance is audited at a somewhat higher granularity, it may still be too replete for comprehension and modification. In particular, when AV techniques are used to create a container, the collected provenance information, being at the file and process level, is still too fine-grained to show the overall workflow.  
	{We consider two kinds of consumers with differing objectives of using the generated provenance graph. Some expert users familiar with execution of the program would like to see the process-view of the provenance graph sans the extraneous low-level system information generated due to auditing of common libraries and system executables. Some users, alternatively, would like to see a summarized graph that is (potentially) closer in appearance to an application workflow or prospective provenance. We  summarize graphs in two ways, in particular collapsing or hiding common libraries and system executables for a process-view, and summarizing retrospective provenance by finding groups of nodes and edges that are related by common ancestry to each other.  Through experiments, we show that our graph summarization methods reduce the actual number of nodes and edges in graphs, by 80--91\%, on average, thus producing meaningful summary graphs.} 
	
	To show use of provenance in sciunits, we first describe how applications can create sciunits using the {\texttt{sciunit}}  
	tool, a Python/C-based Git-like client that creates, stores, and repeats sciunits (Section~\ref{sec:architecture}). Our previous work~\cite{Ton:2017:Sciunit} describes how the {\texttt{Sciunit}} tool uses application virtualization to create containers and store multiple containers in a single sciunit using content de-duplication. In~this paper, we show how to use embedded provenance for exact, partial, and modified repeatability. In~particular, we focus on how provenance associated with two reference executions is matched for exact, partial, and modified repeatability, and how provenance associated with past reference executions is summarized. 
	
	The rest of the paper is organized as follows: Section~\ref{sec:research object evolution} describes related work concerning the evolution of static and reusable research objects and how they utilize provenance. Section~\ref{sec:architecture} describes the \texttt{Sciunit}---its use in applications to build containers and repeat them in various ways, and the embedded provenance model. 
	Section~\ref{subsec:reuse} describes how to utilize embedded provenance for reuse---exact, partial, and modified reuse. Methods for summarizing retrospective provenance are described in Section~\ref{subsec:graph visualization}. Section~\ref{sec:experiments} presents our experiments. Our conclusions and future work are discussed in Section~\ref{sec:conclusion}.

	\section{Related Work}\label{sec:research object evolution}
	
	In this section, we trace the evolution of research objects and how provenance is managed within different kinds of research objects.
	
	{Research objects are increasingly seen as the new social object for advancing science~\cite{DeRoure:DPRM:CRO}. They~are used for dissemination of scholarly work, measuring research impact, and assessing credit and attribution~\cite{Yale:CiSE:2010}, which in the past was mostly done through research papers. The Research Object Model~\cite{Belhajjame:RO:2015, Bechhofer:Linked:2013} is a comprehensive standard defining the concept of a research object as a bundle of  artifacts, specifying a complete digital record of a piece of research. Implementations of the standard have primarily focused on structured workflow objects~\cite{Corcho:Workflow:2012, DeRoure:Preservation:2011, santana2015towards}, and only recently have been extended for general applications (i.e.,  applications executed without a formal workflow system).} 
	
	To create a research object (RO) for a general application, digital artifacts must be placed within it, either manually with explicit commands or automatically by using AV. The former method is used in RO-Manager~\cite{RO-Manager}, 
a tool that uses the RO-Bundle specification~\cite{RO-Bundle}. A more recent approach relies on user action to create the topology, relationship, and node specifications based on a standard~\cite{standard2013topology} that are eventually translated to a container~\cite{qasha2016framework}.  In this paper, we focus on automatically creating research objects using AV.
	
	{Application virtualization is a generic approach to build research objects without modifying applications, and predominantly uses the {\emph{ptrace}} 
	 or {\emph{strace}} system call to create containers~\cite{Guo:2011:CDEFull, Guo:2011:CDEShort}. Some~prominent tools that use AV to build a research object are Sciunit~\cite{Ton:2017:Sciunit, Pham:2013:PTU}, Reprozip~\cite{Chirigati:2013:ReproZip}, Care~\cite{Janin:CARE}, and Parrot~\cite{Thain:2015:Parrot}. Based on AV, all of these tools use \emph{ptrace} to create a manifest of identified dependencies during application run-time. However, application re-execution differs considerably. In Sciunit~\cite{Ton:2017:Sciunit} and Care~\cite{Janin:CARE}, \emph{ptrace} is also used during re-execution time to intercept system calls and redirect them within a native container. This is unlike  Parrot~\cite{Thain:2015:Parrot} and Reprozip~\cite{Chirigati:2013:ReproZip}, which copy dependencies from the manifest into a \emph{chroot} environment or Docker~\cite{Docker} or Vagrant~\cite{Vagrant} container for re-execution.} 
	
	{There are several advantages of using \emph{ptrace} during re-execution time. A native container~is available for instant reuse; if dependencies \textls[-10]{specified in the manifest have to be first copied within another container such as a Docker container, availability of a container for reuse is delayed.~Experimental }results show that redirecting system calls within a container leads to faster re-execution times~\cite{meng2015invariant}. However,~more importantly, using \emph{ptrace} enables transparent provenance auditing of the application~\cite{Pham:2013:PTU} both during container creation and re-execution time. Care~\cite{Janin:CARE} does not audit provenance at either container creation or re-execution time. Therefore, in this paper, we have used \HT{\texttt{Sciunit}~\cite{SciunitI:2017:sciunit.run}} for understanding how to manage provenance in research objects. In this paper, we show how the audited provenance can be used for reusing research objects, in particular to verify correctness of results.}
	
	{Using \HT{\texttt{Sciunit}~\cite{SciunitI:2017:sciunit.run}} does not limit the use of commercial container technologies such as \HT{Docker~\cite{Docker} and Vagrant~\cite{Vagrant}}. In Sciunit, commercial containers such as Docker are merely a wrapper for standardization, since application virtualization creates a self-contained container, and the translation to Docker files from the collected dependency information is fairly straightforward. Another advantage is that Sciunit versions the contents of the containers using content de-duplication techniques~\cite{Muthitacharoen:ContentBaseddeDuplication} and thus it can produce a versioned provenance graph.}
	
	{Transparent provenance auditing can be achieved at different granularities. In NoWorkflow~\cite{murta2014noworkflow}, it is done at the level of the abstract syntax tree of Python programs. In Sciunit, it is independent of the programming language and at the system level. Furthermore, to support efficient containerization of application programs, Sciunit differs from PASS~\cite{muniswamy2006provenance} and \HT{SPADE~\cite{gehani2012spade, malik2013sketching}}. In particular, Sciunit audits at the granularity of file open and close and PASS and \HT{SPADE} audit at the granularity of file reads and writes. \HT{The provenance} model in Sciunit \HT{has also been extended} to include database applications~\cite{Pham:ICDE:LDV} and distributed applications~\cite{Pham:2014:Thesis}.}
	

	
	
	{\textls[-5]{Understanding application execution within a research object can incentivize its re-use.~Provenance} audited with application virtualization methods generates fine-grained provenance, not useful for user consumption. Methods that link retrospective provenance to prospective provenance~\cite{Dey:2015:LinkingProReTro} are useful, except that, in most reusable research objects, there is no formal guarantee that application workflow in the form of prospective provenance shall necessarily be available. In addition, Sciunit creates containers independent of programming languages. Thus, assumptions such as users annotating source code of script files as in YesWorkflow~\cite{YesWorkflow} cannot be made. Therefore, we focus on methods that summarize provenance without assuming the description of prospective provenance.}
	
	{Several methods for provenance graph summarization have been proposed~\cite{MMS13, Tian:2008:SNAP, CCBD06}. We classify them as statistical~\cite{MMS13} and non-statistical~\cite{Tian:2008:SNAP, CCBD06} methods.  Statistical methods use techniques such as clustering or user-defined views to determine relevant nodes. Non-statistical methods are based on pure aggregation of nodes and derivation histories. In our experience, non-statistical methods are easier to implement and \HT{can be} used in light-weight databases, such as LevelDB, embedded within containers than clustering based methods, which assume presence of a graph or relational database. Therefore, in~this~paper, we have focused on non-statistical methods. \HT{Within} non-statistical methods, we focus on spatial summarization, i.e., reducing the number of nodes and edges in a single provenance graph and not temporal summarization i.e., summarization across multiple provenance graphs as considered in SGProv~\cite{Daniele:2014:SGProv}. This is because the first objective is to understand application execution and therefore the objective of summarization is to generate a summary as close as possible to prospective provenance. Temporal summarization can also be useful as users compare initial application workflow with its re-runs but is currently beyond the scope of this paper.}

	\section{Using \texttt{Sciunit}}
	\label{sec:architecture}
	
	We describe how the \texttt{Sciunit} \HT{client} is used to create reusable research objects. By design, sciunit is both the name of the reusable research object we define and the name of the command-line \HT{client}. The~\texttt{Sciunit} \HT{client} creates, manages, and shares sciunits. 
	
	\subsection{A Sample Application}\label{subsec:usecase}
\begin{figure}[H]
		\centering
		\includegraphics[width =14.5cm]{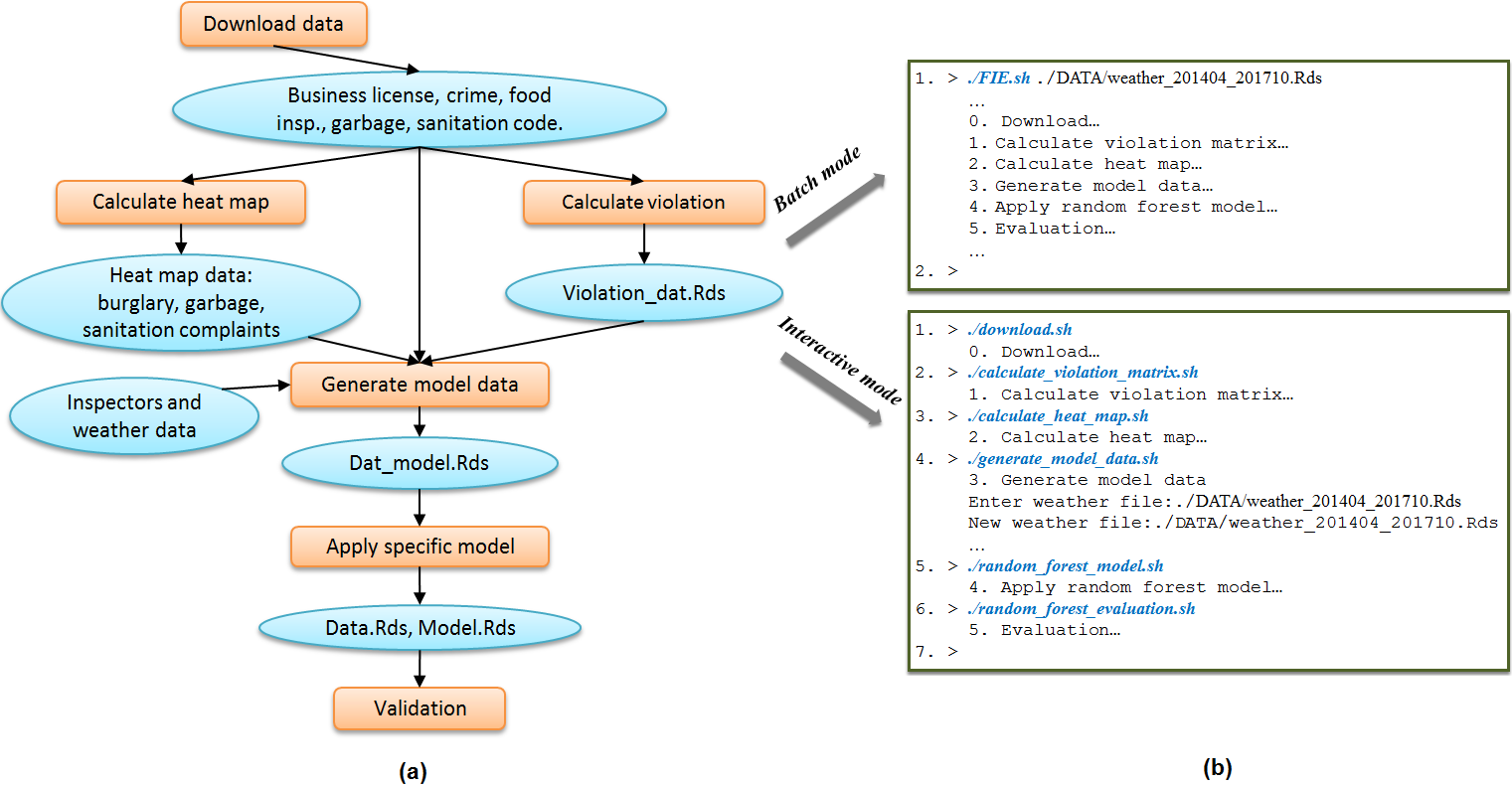}
		\caption{(\textbf{a}) \HT{C}onceptual view of the steps required to run the Food Inspection Evaluation~\cite{FIE} predictive model; (\textbf{b}) \HT{T}wo possible execution modes.}
		\label{fig:FIE workflow}
	\end{figure}

	Our reference implementation is the \texttt{sciunit}, a Python/C command-line client program that creates reusable research objects, stores them efficiently, and repeats and reproduces them~\cite{SciunitI:2017:sciunit.run}. To~demonstrate the primary commands and salient features of the client program, we use a real-world example. Figure~\ref{fig:FIE workflow}\HT{(a)} shows an example of a predictive model used for forecasting critical violations during sanitation inspection, known as Food Inspection Evaluation (FIE)~\cite{FIE}. The software consists of scripts written in different languages (R and Shell) that operate on input datasets acquired from the City of Chicago Socrata data portal~\cite{CoC}. The output of the predictive model is continually tested using a double-blind retrodiction; the Department of Public Health conducts inspections via its normal operational procedure, which are compared with the output of the model. The  pre-processing code is shared on GitHub.
	\cite{FIE-C},  the data is available via public repositories~\cite{CoC}, and the predictive model analysis is also published~\cite{FIE-P}. Bundling these artifacts into a mere shared research object would likely be inefficient given data from nine different sources, which changes periodically, making analysis conducted within a certain time range obsolete. A reusable research object is needed. 
	
	\subsection{Creating, Storing, and Repeating a Container with \texttt{Sciunit}}\label{subsec:creatingAContainer}
	The FIE predictive model can be run in two modes, either as a batch mode, using a Shell script that serially executes all sub-tasks 
	or in an interactive mode, wherein the user provides some input parameters to few sub-tasks such as weather files in a specific date range. Figure~\ref{fig:FIE workflow}\HT{(b)} shows the {two possible executions}. The \HT{\texttt{Sciunit}} client can be used to build a reusable research object consisting of identifiers of one or more re-executable containers in both the batch and interactive modes.
	
\begin{figure}[H]
		\centering
		\includegraphics[width = 15cm]{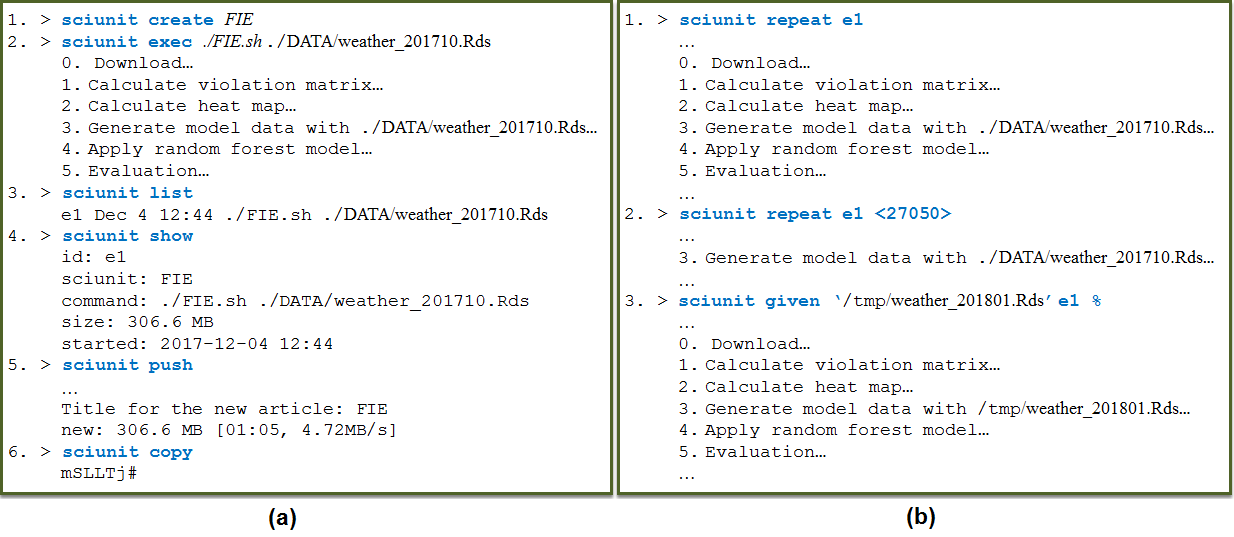}
		\vspace{-3pt}
		\caption{User interaction with the Sciunit client: (\textbf{a}) audit mode and (\textbf{b}) repeat mode.}
		\label{fig:client interaction}
	\end{figure}

	Figure~\ref{fig:client interaction}\HT{(a)} shows a sample user interaction with \HT{\texttt{Sciunit}} client for auditing the FIE program. The~user creates a namespace sciunit titled \emph{FIE} (Line\,1). To create a container within the sciunit, the~user runs the application with the \emph{exec} command (Line\,2). Packaging an application into a container also audits provenance information of the application run. Many containers, each corresponding to a given execution of the client program, can be created within the same \emph{FIE} sciunit by using the \emph{exec} command again. All executions can be listed with the \emph{list} command (Line~3) and the last execution can be listed with the \emph{show} command (Line~4).
	
	\pagebreak
	
	\textls[-15]{The \emph{exec} command makes minimal assumptions regarding the nature of the application. In~particular, }the user application can be written in any combination of programming languages, e.g., C, C++, Fortran, Shell, Java, R, Python, Julia, etc. or be used as part of a workflow system such as Galaxy~\cite{Goecks:2010:Galaxy}, Swift~\cite{Zhao:2007:Swift}, Kepler~\cite{Altintas:2006:Kepler}, etc. While our description assumes local execution, in practice, an application's execution can be either local or distributed. {We choose an example with local execution since the AV methods for distributed and parallel applications are currently not integrated with \HT{\texttt{Sciunit}}, and cannot generate the required provenance graph. An AV method for database applications is outlined in Light-weight Database Virtualization (LDV)~\cite{Pham:ICDE:LDV} and for high performance computing (HPC) programs in \HT{Pham Q.'s thesis~\cite{Pham:2014:Thesis}.}} 
	
	%
		
	The created FIE sciunit and associated containers are stored locally unless explicitly shared with a remote repository using the \emph{push} command, which instructs the client to upload \HT{the sciunit and} all the containers in a sciunit to a Web-based repository (see {Line\,4} in Figure~\ref{fig:client interaction}\HT{(a)}). The \HT{\texttt{Sciunit}} client uses Hydroshare~\cite{Hydroshare} for geoscience applications and Figshare~\cite{Figshare} otherwise as its Web-based repository. The \HT{\texttt{Sciunit}} client also supports \HT{sharing} with \emph{copy} command ({Line~8}, Figure~\ref{fig:client interaction}\HT{(a)}). In order to copy a \HT{sciunit} to \textit{client2}, \textit{client1} should have the {\textbf{<tokenID>}}  
	generated by the command \textit{sciunit copy} and used by \textit{client2} to open the sciunit (i.e., \emph{sciunit open <tokenID>}). The \HT{sciunit} is transferred from \textit{client1} to \textit{client2} through a \HT{third-party} cloud-based web service. 
	
	{A container within a sciunit (identified by an increasing sequence) can be re-run on the local machine with the \emph{repeat} command. Users can either exactly repeat  the entire computation by calling \emph{repeat} with execution ID (see \HT{L}ine~1, Figure~\ref{fig:client interaction}\HT{(b)}) or partially repeat some processes in this computation by giving a list of processes ID they want to repeat (see \HT{L}ine~2, Figure~\ref{fig:client interaction}\HT{(b)}).}
	
	{The option to modify data inputs or program files is also available in sciunit with the \emph{given} command. This functionality allows users to re-execute the packages with their own local data inputs or new program files that may be stored outside container. For instance, in our example (see \HT{L}ine~3 in Figure~\ref{fig:client interaction}\HT{(b)}), the \emph{FIE} program is repeated with the new data input (i.e.,  ``/tmp/weather\_201810.Rds'') at a local directory.}

	{A sciunit may include many containers, each container corresponding to one reference execution. Each time an application is audited, duplicate file dependencies of the application can be copied into the sciunit. To avoid redundancy, \texttt{Sciunit} checks for duplicate dependencies as the container is created during the AV audit phase. \texttt{Sciunit} uses content-defined chunking to divide the container's content into small chunks identified by a hash value, as described in detail in our prior work~\cite{Ton:2017:Sciunit}.}
	
	
	\section{Reusing Sciunits}\label{subsec:reuse}
	
	{Sciunit distinguishes between an \emph{execution trace}  and a \emph{provenance dependency trace}. \HT{\emph{Ptracing}} an application generates an execution trace, which is a \HT{log} of the execution of activities in the container. However, it does not generate the correct \HT{causality or } dependency information leading to a provenance~trace. \HT{In other words, connectivity in the log does not necessarily imply dependency. Consider,  the simple execution trace in Figure~\ref{fig:dependency} as logged with temporal annotations of when $P_1$, and $P_2$ used and wrote to files $A$, $B$, and $C$. If we consider only the edges of the execution trace there exists a path between $A$ and $C$. However, $C$ cannot depend on~$A$ due to temporal constraints. This is because $P_2$ stopped reading $B$ before it was written by $P_1$.}
	
	\HT{We consider a simple inference algorithm to determine a provenance dependency trace by determining the state of a node in the extution trace. More formally, an execution trace is a labeled directed graph $\exTrace = (V,E,T)$  with~nodes $V$ and edges $E \subseteq V \times V$. Each node must be of one of the activity and entity~types, wherein an activity corresponds to a process and an entity corresponds to a~file. Each~edge} with an allowed start and end activity or entity type has a label from ${\mathcal{L}}=  \{readFrom(file,process), hasWritten(process,file), executed(process,process)\}$, and a function $T: E \to \mathbb{T}
		\times \mathbb{T}$, mapping edges to intervals from a
		discrete time domain $\mathbb{T}$. We use $T(v_1,v_2)$ to denote the
		time interval associated by $T$ to the edge $(v_1, v_2)$ and
		$\intBegin{I}$ and $\intEnd{I}$ to denote the lower respective upper bound
		of an interval $I$. Thus, each edge is annotated with a time interval indicating when the two connected nodes interacted: for example, the time interval during which a process (activity) was reading from a file (entity), or a time at which a process forked another process. A simple inference algorithm to determine a provenance dependency trace is by determining the state of a node in $\exTrace$. The state of an entity $e$ depends on an entity $e'$ at a time $T$ if (i) there is a path between $e'$ and $e$ in the execution trace; and (ii)  temporal annotations on the edges of the path do not violate temporal causality. That is, there exists a sequence of times $T_1, \ldots, T_n$ so that for each path we have $T_i \leq T_i + 1$ and $T_i \leq T(v_i, v_i+1)_e$. In other words, the information flows from and entity $e_2$ to $e_1$ complies with the temporal annotations.
		
		\HT{We further assume that provenance execution trace has no cycles, since repeat execution is not guaranteed if the trace has cycles. 
Consider a simple example in which two processes $P_1$ and $P_2$ access file $F_1$. $P_1$ runs at time $t_1$ and reads from $F_1$. After that, $P_2$ runs at time $t_2$ and writes to $F_1$ ($t_1 < t_2$). Since $F_1$ is accessed by both $P_1$ and $P_2$, it will be included in the container. However, since the content of $F_1$ was modified at $t_2$ by process $P_2$, process $P_1$ cannot be exactly repeated as the first time it was run. 	
{This problem can be avoided if the file $F_1$ is versioned. Suppose the container has the capability to version the resources/dependencies, and the container will keep two  versions of $F_1$: $F_1^1$ and $F_1^2$ with respect to $F_1$ before and after $t_2$. At the repeating time, $P_1$ {will be fed} with the $F_1^1$ keeping the original data of $F_1$. Through this method, $P_1$ will read data exactly as the first time it was executed. Experiments show that versioning each file has an overhead. However, in~\texttt{Sciunit}, it can be enabled for special cases such as when auditing a concurrent program.}}\
	
		\begin{figure}[H]
		\centering
		\includegraphics[width = 3.5 in]{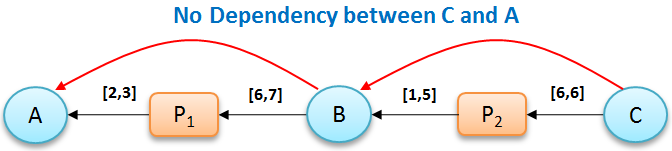}
		\caption{An example of no dependency.}
		\label{fig:dependency}
	\end{figure}

	{Given a valid provenance dependency information with no cycles, the \texttt{Sciunit} can use this  to enable the various commands shown earlier. In particular, \texttt{Sciunit} can use the provenance graph to (i) simply repeat the container exactly as shared; (ii) repeat some identifiable part of the application flow; and (iii) repeat but with different input arguments, producing a different but valid output. We term them \emph{exact}, \emph{partial}, and \emph{modified} repeat executions. During exact repeat execution, provenance is used for verifying if the execution was repeated exactly as the previous reference execution. During partial repeat execution, provenance is used to build a subset container containing the necessary and sufficient dependencies to run the part of the application flow. During modified repeat, provenance is used to establish which part of the container can be re-used. We describe these operations in more detail.}

	

	
	\subsection{Exact Repeat Execution}\label{subsubsec:exact}
	{Exact repeat execution refers to the process of running a computation again (usually on a different environment) with the same inputs and obtaining the same outputs. A container within a sciunit (identified by an increasing sequence) can be re-run exactly on the local machine with the \emph{repeat} command (Figure~\ref{fig:client interaction}b (Line 1)). To verify if repeat produced exactly the same outputs, the generated entities must be hashed and they must be produced in exactly the same way as they were in the reference execution. }
	
	{In \texttt{Sciunit}, content validation is done through the versioning system that de-duplicates content. Even if the versioning system validates the same content, some temporary output files may have different names, and labels of processes such as its ID are not guaranteed to be identical every time application re-executed. To measure the correctness of repeatability, we focus our effort on comparing provenance graphs through their node structure. Since the provenance graph records all information about the execution, having exact repeat execution means the provenance graph included in the container at audit time and new provenance graph generated during repeat execution are isomorphic.}
	
	{To begin with, we first define the term provenance isomorphism as follows:}
			
	{\begin{defi}[\textbf{{Provenance Isomorphism}}] \label{def:provenanceIsomorphim} \emph{Given a set of nodes $V=\{Activity(V_A), Entity(V_E)\}$ and a set of edge labels $\mathcal{L}=\{L_1, L_2, ..., L_n\}$, two provenance graphs $G=\{V, E_G\}$ and $H~=~\{V, E_H\}$ are said to be isomorphic if there is a non-trivial automorphism, i.e., a bijective function }
			\begin{math}
			f:\left\{
			\begin{array}{l}
			Activity(V_A) \longrightarrow Activity(V_A)\\
			Entity(V_E) \longrightarrow Entity(V_E)
			\end{array}
			\right.
			\end{math}
			\emph{such that $e_h=\{u, v\} \in E_H$, $Type(e_h) = L$ if and only if $e_g~=~\{f(u), f(v)\} \in E_G$, $Type(e_g)~=~L$.}
	\end{defi}}
	
	\HT{In particular, provenance graphs are labeled with nodes labeled as activity nodes or entity nodes referring to processes and files respectively and edges labeled based on types used in W3C PROV standard: $\mathcal{L}=\{used, wasGeneratedBy, wasInformedBy\}$.} Many other algorithms such as Nauty~\cite{McKay:1981:Nauty, McKay:2017:Nauty}, i.e.,  considered as the fastest general graph isomorphism algorithm 
	search for the whole automorphism group (all isomorphism bijections between two graphs). This is computationally hard and can lead to longer execution times. Meanwhile, our algorithm, i.e.,  applied for provenance isomorphism (a special kind of graph isomorphism) as defined in Definition~\ref{def:provenanceIsomorphim}, is polynomial, since having at least one bijective function is enough to claim two provenance graphs are isomorphic. We find this one bijective function by comparing the node hashes computed by taking into account its neighbors.} 
	
	{Our Algorithm~\ref{algorithm:similar graph} describes the details of provenance isomorphism verification process. Given two input provenance graphs (i.e., $G_1$ and $G_2$), Algorithm~\ref{algorithm:similar graph} outputs a bijective function (i.e., $f: R_1 \longrightarrow R_2$) if these two graphs are isomorphic. Otherwise, it returns $False$ (Line 7). The first step of this algorithm is to calculate the $HashValues$ for each node in each graph by using function {\textbf{buildHashValues}} 
	 (Lines~5--6). Particularly for each node $u$ in graph $G$, this function concatenates all its edge types and its neighbor labels to its $HashValues$ (Lines 9--11). Next, it turns to find a bijective function by calling \textbf{findBijection()} (Line 7). This function sequentially takes a node $u_1^i$ in $G_1$ and considers each candidate $u_2^i$ in $G_2$. If these two nodes both have the same type and similar $Hashvalues$ (Lines~15--17), then it recursively continues to go further with smaller graphs (Lines 18--20) until it finds a bijective function when $G_1$ is empty (Line 27). Otherwise, it considers other candidates in $G_2$ (Lines 23--25). It~may also turn to $False$, if no candidate in $G_2$ is found (Line 26).}

\pagebreak

	\begin{algorithm}[H]		        
		\DontPrintSemicolon
		\SetKwProg{Def}{}{:}{} 
		\SetKwInOut{Input}{Input}
		\SetKwInOut{Output}{Output}               
		\Def{\textbf{ProvenanceIsomorphism} (
		$G_1$, $G_2$, $R_1$, $R_2$)} {
			\Input{two provenance graphs $G_1$ and $G_2$}
			\Output {a bijective function $f: R_1 \longrightarrow R_2$}
			$R_1 = R_2 = Empty$\;
			\If{(($G_1$ is Empty) || ($G_2$ is Empty))}{
				\textbf{return} False\;
			}
			\textbf{buildHashValues} ($G_1$) /* Add hash values for each node in graph */\;
			\textbf{buildHashValues} ($G_2$)\;
			\textbf{return} \textbf{findBijection} ($G_1$, $G_2$, $R_1$, $R_2$) /* Find a bijection between two graphs */\;
		}
		\Def{\textbf{buildHashValues} ($G$)} {			
			\ForEach {node $u$ in $G$}{
				\ForEach{edge $e = \{(u, v)~or~(v, u)\}$ connects to node $u$}{
					Add $\{Type (e), Label (v)\}$$\quad$to$\quad$$u.HashValues$\;
				}
			}
		}
		\Def{\textbf{findBijection} ($G_1$, $G_2$, $R_1$, $R_2$)} {
			\If{($G_1.length$ 
			$\#$ $G_2.length$)}{
				\textbf{return} False\;
			}
			\ForEach{node $u_1^i$ in $G_1$}{
				\ForEach{node $u_2^i$ in $G_2$}{
					\If {($(Type(u_1^i)==Type(u_2^i))$ $\&\&$ ($u_1^i.HashValues$ and $u_2^i.HashValues$ are similar))}{
						Remove $u_1^i$ from $G_1$ and push $u_1^i$ to $R_1$\;
						Remove $u_2^i$ from $G_2$ and push $u_2^i$ to $R_2$\;
						\If {findBijection ($G_1$, $G_2$, $R_1$, $R_2$)}{
							\textbf{return} True\;
						}
						\Else{ 
							Pop $u_1^i$ from $R_1$ and add $u_1^i$ to $G_1$\;
							Pop $u_2^i$ from $R_2$ and add $u_2^i$ to $G_2$\;
							\textbf{continue}\;
						}
					}
				}
				\textbf{return} False\;
			}
			\textbf{return} True\;
		}
		\caption{Checking the exact execution using provenance graphs}
		\label{algorithm:similar graph}
	\end{algorithm}


\subsection{Partial Repeat Execution}\label{subsubsec:PartialRepeat}
	
	To partially repeat, a user selects one or multiple processes within a container. These processes are identified by their short pathname, or PID, and the user can also use the provenance graph to aid in identification. While the provenance graph can be quite detailed for a user to choose specific processes, in Section~\ref{subsec:graph visualization}, we describe how a user can see a process view or a summarized application workflow akin to the workflow presented in Figure~\ref{fig:FIE workflow}a from the provenance graph. Thus, for example, using the container from Figure~\ref{fig:FIE workflow}, a user selects the processes ``Calculate violation'' and ``Generate model data'' as the group of processes to be partially repeated. Since this user-selected group of processes may not include all related processes needed for re-execution, we must determine these related processes, along with the data files they reference. The determined processes and files will constitute the new ``partial repeat'' container or ``sub-container''. Algorithm~\ref{algorithm:build sub-container} shows the procedure for building the sub-container. It starts with the list of user-selected processes (\emph{selectedProcs}), and progresses to include all relevant processes and files by traversing the lineage of the graph {(Lines~10--24)}. The \emph{getDeps} function assumes that any intermediate data files, if included as dependencies, still exist as generated from previous execution runs. {The \emph{isDirectDecendant} function is used to detect \HT{direct} decendant processes that need to be included. Meanwhile, the \emph{directResources} function marks all data files and dependencies \HT{directly} touched by any process in \emph{requiredProcs}}. The execution of this algorithm ensures that the data file ``Heat map data'' generated from the previous run of the process ``Calculate heat map''  is included in the sub-container, even though, in the new partial repeat execution, the process ``Calculate heat map''  will not be re-executed.\\

	\begin{algorithm}[H]		        
		\DontPrintSemicolon
		\SetKwProg{Def}{}{:}{}                
		\Def{\textbf{BuildSubContainer} (
		$selectedProcs$, $container$)} {
			$subContainer$ = initialize ($container$)\;
			$allProcs$ = getAllProcs ($container$)\;
			$requiredProcs$ = \textbf{getProcs} ($selectedProcs$, $allProcs$)\;
			$reqProcDeps$ = \textbf{getDeps} ($requiredProcs$)\;
			\ForEach{$dep$ \textnormal{\textbf{in}} \{$reqProcDeps$\}} {
				/* add dep to correct location in subContainer */\;
				add ($dep$, $container$, $subContainer$)\;
			}
			\textbf{return} $subContainer$\;
		}
		\Def{\textbf{getProcs} ($selectedProcs$, $allProcs$)} {
			$result$ = \{$selectedProcs$\}\;
			\ForEach{$proc$ \textnormal{\textbf{in}} \{$allProcs$\}} {
				\ForEach{$selProc$ \textnormal{\textbf{in}} \{$selectedProcs$\}} {
					\If{\textnormal{\HT{isDirectDecendant}} ($proc$, $selProc$)} {
						$result$ = $result$ $\cup$ $proc$\;
						\textbf{break}\;
					}
				}				
			}
			\textbf{return} $result$\;
		}
		\Def{\textbf{getDeps} ($requiredProcs$)} {
			$result$ = $\emptyset$\;
			\ForEach{$reqProc$ \textnormal{\textbf{in}} \{$requiredProcs$\}} {
				/* retrieve all related files and dependencies */\;
				\HT{$deps$ = directResources ($reqProc$); /* get all the directly accessed files or dependencies */} \\
				$result$ = $result$ $\cup$ $deps$\;				
			}
			\textbf{return} $result$\;
		}
		\caption{Build sub-container for partial execution}
		\label{algorithm:build sub-container}
	\end{algorithm}
	\vspace{12pt}

	\subsection{Modified Repeat Execution}\label{subsubsec:ModifiedRepeat}
	{In repeating an execution exactly, a computation is repeated with the same inputs, and obtaining the same outputs. Modified repeat execution refers to the notion of reproducibility. Reproducibility refers to the process of running a computation again with different inputs and observing the outputs. The outputs of a reproduced computation may be checked against \emph{expected} outputs to validate the logic of the computation. Alternately, a computation may also be reproduced by altering the computational logic itself. In reproducibility, expected outputs are user-defined and, in general, hard to verify. Provenance can still be useful for modified repeat execution. }
		
\begin{figure}[H]
		\centering
		\includegraphics[width = 1 in]{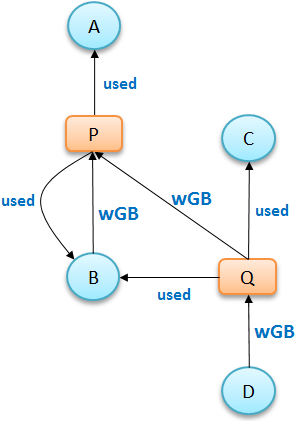}
		\caption{Provenance graph of an execution.}
		\label{fig:VersioningExample}
	\end{figure}

	{Consider the provenance graph of an execution in Figure~\ref{fig:VersioningExample}. This execution consists of two processes: $P$ and $Q$. Files $A$ and $C$ are used by $P$ and $Q$, respectively. $B$ is an 
		`intermediate' input produced by $P$ and used by $Q$, which itself is spawned by $P$, and uses $B$ and $C$ as inputs to produce final output $D$. Repeating this execution would entail running it again with the exact same inputs (i.e.,  `unchanged' data files) for $A$ and $C$, at which the exact same result for output $D$  will be produced. Reproducing this execution using the Sciunit \emph{given} command implies running with a modified inputs, either $A$ or $C$, or both.} 
		
	{If $A$ or both inputs are modified, the entire execution must be re-run again. However, if only $C$ is modified, then the only part of the computation that will run differently is $Q$ (i.e.,  $P$ will produce the same output $B$, given the same input $A$). If $P$ is far more time-consuming than $Q$, it might suffice to run only $Q$, avoiding the expense of running $P$ again. Reducing the unneeded processing time is often critical if the execution is to be altered for a large number of modifications to input $C$. This~partial reproduction would be possible if, upon repeating the computation with its original inputs, the~intermediate output $B$ produced by $P$ was saved in the container. }
	
	{We use the embedded provenance graph to determine which part of the provenance graph need not be reprocessed again. Our algorithm is the same as Algorithm~\ref{algorithm:build sub-container} in that we identify the primary processes of changed inputs, and from that determine the necessary and sufficient dependencies (i.e.,  \emph{getDeps} function). This is the part of the graph that must be re-run. Nodes that are not in this dependency set are simply re-used from the container. }


	%
	
	\section{Summarizing Provenance Graphs}\label{subsec:graph visualization}
	
\begin{figure}[H]
		\centering
		\includegraphics[width = 13cm]{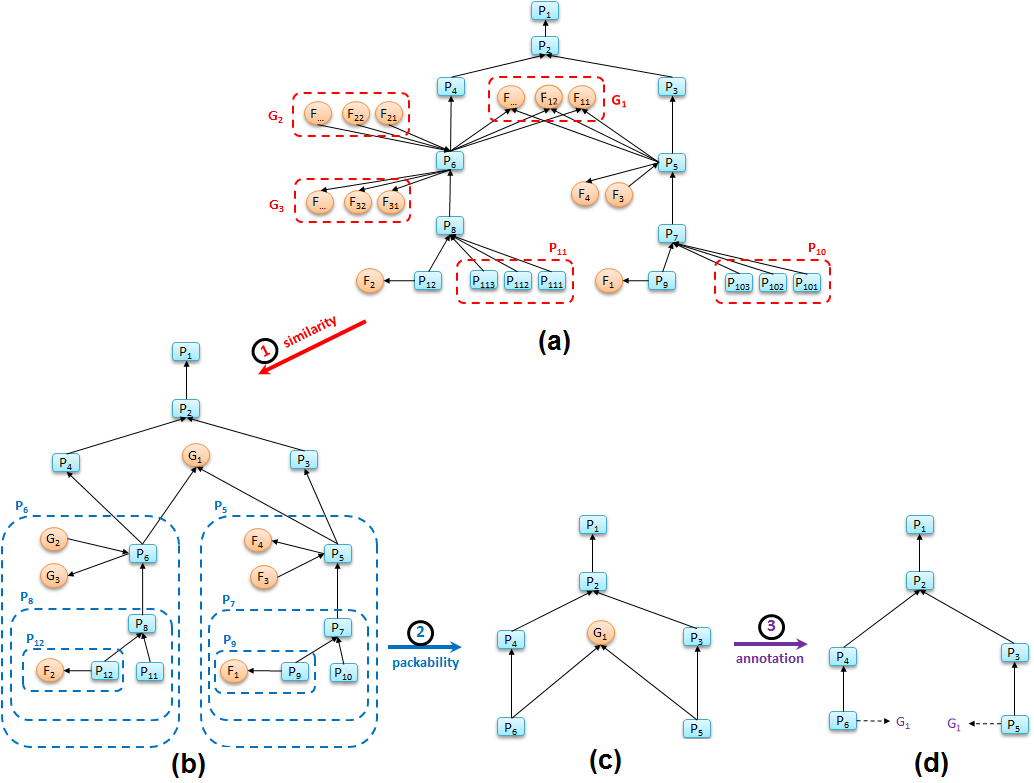}
		\caption{Graph summarization of a replete graph: (a) orginial provenance graph, (b) applying similarity rule, (c) applying packability rule and (d) final sumary graph.} 
		\label{fig:Graph Optimization}
	\end{figure}

	Provenance information generated by AV audit methods is fine-grained. A graph created~from a complete set of generated provenance, \textls[-5]{using normal visualization structures such as tree or list representations, would be far too replete to be of real practical value. When viewed, this graph~would }present significant system-level detail that would inhibit a basic comprehension of the overall application workflow. For example, the intuitive workflow of Figure~\ref{fig:FIE workflow}\HT{(a)}, consisting of 12 nodes and 13 edges, \HT{is} represented fully as a dense \HT{provenance} graph of 146 nodes and 321 edges. Figure \ref{fig:Graph Optimization}\HT{(a)} shows a part of this replete graph redrawn for visual clarity. 
	
	The definition of `intuitive' is subjective. We consider two \HT{use cases} with differing objectives in using the generated provenance graph: {(i) a predominant process-view of the provenance graph sans the extraneous low-level system information generated due to auditing of common libraries and system executables; and (ii) a summarized graph that is (potentially) closer in appearance to an application workflow or prospective provenance as in Figure~\ref{fig:FIE workflow}\HT{(a)}. In this section, we describe two formal methods for (i) and (ii). In (i), the key idea is to collapse or hide as much as possible common libraries and system executables, thus summarizing retrospective provenance; and, in (ii),  the key idea is to summarize by finding groups of nodes and edges that may be related semantically to each other so as to potentially match with prospective provenance.  We evaluate the first method based on total number of system information collapsed and second method based on available \HT{Unified Modeling language (UML)} diagrams of our sample test programs. However, UML diagrams are not assumed as inputs to the summarization~method \HT{and present as part of the sciunit.}}

	\subsection{Collapsing Retrospective Provenance}
	
	{Given a directed graph $G=(V,E)$, where $V$ is the set of vertices {(in our graph, a vertex is of type ``file''  or of type ``process'')} and $E$ is the set of edges, we denote $Input(u)$ and $Output(u)$ as the sets of input and output edges of vertex $u$. Respectively, $Input(u)=\{e|~\exists v \in V,~e=(v,u) \in E \}$, and $Output(u)=\{e|~\exists v \in V,~e=(u,v) \in E\}$. The direction of an edge characterizes the dependency of its vertices.  For example, a process $u$ spawned by process $v$ is represented by the edge $(u,v)$, and a file $u$ read by process $v$ is represented by the edge $(v,u)$. The graph $G$ is collapsed based on the following two rules:}

	\begin{Def}{\textbf{Similarity.}}
		\emph{Two vertices $u$ and $v$ are} 
		 \emph{called \textit{similar} if and only if they share the same type and have the same input and output connection sets: $Type(u)=Type(v)$, $input(u)=input(v)$ and $output(u)=output(v)$}.
	\end{Def}
	
	The similarity rule groups multiple vertices into a single vertex if the vertices \HT{(i)} have the same type and \HT{(ii)} are connected by the same number and type of edges. Additionally, edges of similar vertices will be grouped into a single corresponding edge. Since the provenance graph follows W3C PROV-DM standard, each file is of type entity and each process is of type activity. When applied to our provenance graph, \HT{this rule groups different files and processes that are similar each other into the summary groups of files and processes (see Figure~\ref{fig:Graph Optimization}(a))}.
	
	\begin{Def}{\textbf{Packability.}}
		\emph{A vertex $u$ belongs to $v$'s $generalization~set$ if and only if vertex $u$ connects to $v$ and satisfies one of the following conditions:}
		\begin{itemize}[leftmargin=*,labelsep=5.8mm]
			\item \emph{Vertex $u$ is a file that has only one edge to process $v$: $Type(u)=file$ and \{$\exists ! e~|~e \in E \wedge (e~=~(u,v) \vee e =(v,u))$\}.}
			\item \emph{Vertex $u$ is a process that has only one output edge to process $v$: $Type(u)=process$ and \{$\exists ! e~|~e \in E \wedge e = (u,v)$\}.}
			\item\emph{ Vertex $u$ is a file that has only two edges---an output edge to process $v$ and an input~{edge from another process}\textls[-20]{ $x$: $Type(u)=file$ and \{$\exists ! (e_1,e_2)~|~(\exists x \in V,~v \neq x)~ \wedge (e_1=(u,v) \in~E,\, e_2~=~(x,u) \in E)$\}}.}
		\end{itemize}
	\end{Def}
	
	The packability rule identifies hubs in the provenance graph by packing files or processes that are connected {by single edges to their parent nodes}. It also packs files that are generated and consumed by a single process into their parent processes by producing a process-to-process edge.
	
	When applied in sequence, the similarity and packability rules condense the detail-level of a graph while preserving its core workflow elements. Figure~\ref{fig:Graph Optimization} (all process names and file names are simplified for brevity) illustrates how applying these two rules to a replete graph produces a graph summary that shows the primary processes in a workflow. Figure~\ref{fig:Graph Optimization}a presents the original replete provenance graph of one sub-task of the FIE workflow (the data processing steps ``Calculate Violation'' and ``Calculate Heat Map'' of Figure~\ref{fig:FIE workflow}\HT{(a)}). Applying the two summarization rules produces the graph in Figure~\ref{fig:Graph Optimization}\HT{(c)}. 
	
	We use an annotation method that assigns higher collapsibility to file nodes than process nodes, since an application workflow is typically defined by the primary processes that it runs.  Figure~\ref{fig:Graph Optimization}\HT{(d)} shows how the annotation {``G\_1''}, which is a library dependency used both by {``P\_5'' and ``P\_6''}, is attached to the two process nodes that generated it. Thus, given a file with $n$ edges ($n \geq 2$), we replace this file with $n$ annotations. 
	
\begin{figure}[H]
		\centering
		\includegraphics[width=5in]{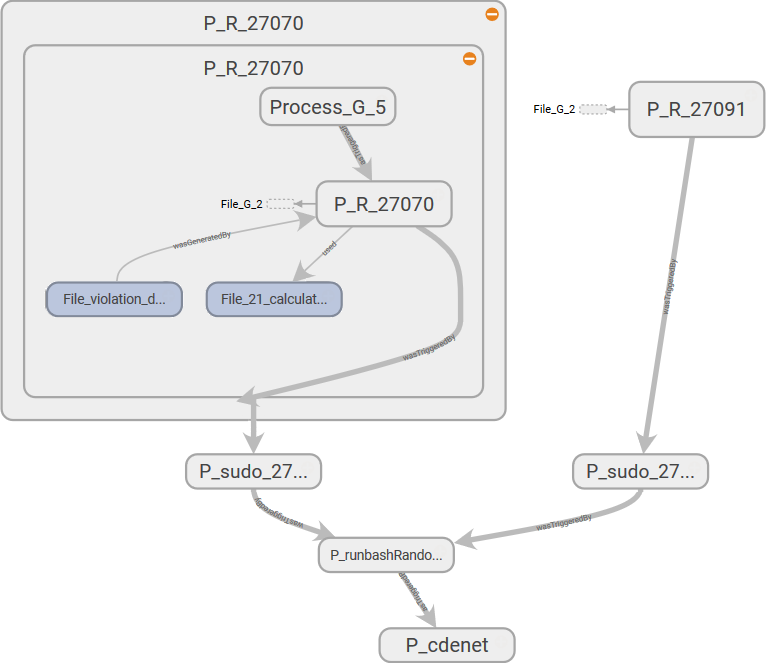}
		\caption{Expanded view of concealing node ``P\_R\_27070'' (``P\_5'').}
		\label{fig:Detailed View}
	\end{figure}

	{Figure~\ref{fig:Detailed View} shows the expanded view of node ``P\_R\_27070'' (``P\_5'' in Figure~\ref{fig:Graph Optimization}\HT{(b)})}. {In Figure~\ref{fig:Detailed View}}, similarity and packability rules group the nodes within the box into the single node {``P\_R\_27070'' (process 27070 runs a subprocess using file ``21\_calulate\_violation\_matrix.R'' (``F\_3'' in Figure~\ref{fig:Graph Optimization}\HT{(b)}) and write data to file ``violation\_data.Rds'' (F\_4 in Figure~\ref{fig:Graph Optimization}\HT{(b)})). These nodes are application nodes and not system nodes. 
		Here, ``Process\_G\_5'' (P\_7 in Figure~\ref{fig:Graph Optimization}\HT{(b)}), another concealing node, correctly hides all the dependencies of the R process calculating the violation matrix.}

	\subsection{Summarizing Retrospective Provenance to Generate Prospective Provenance}
	\label{subsec:efficient graph optimization}
	The method in the previous section summarizes retrospective provenance by collapsing information. However, such summaries still differ from the conceptual view of the applications. For example, Figure~\ref{fig:FIE workflow}\HT{(a)}, which is more familiar to users, is very different in view than Figure \ref{fig:Graph Optimization}\HT{(d)}. Another equally important goal of summarization is to summarize retrospective provenance such that it (potentially) matches application workflow. In some situations, this application workflow may be available in the form of prospective provenance~\cite{Dey:2015:LinkingProReTro}. If available, summary methods can take advantage of this available information. In containers created of ad hoc applications, however, application workflows are rarely available. Therefore, we describe a summarization method that determines the lineage history of nodes and uses this information to summarize retrospective provenance. 
	

	Our method to summarize retrospective provenance is based on ideas described in SNAP (Summarization by Grouping Nodes on
		Attributes and Pairwise Relationships)~\cite{Tian:2008:SNAP}. {SNAP is a non-statistical method for summarizing
undirected and directed graph nodes and edges based on their respective types. In brief, it first groups nodes of the same type. It then recursively sub-divides to form smaller groups of nodes that still have have the same node type but also same relationship type with other groups. SNAP considers direct relationship types amongst group nodes and not relationship types due to ancestry of the nodes. Thus, grouping provided by SNAP can be further improved for provenance graphs by considering ancestral history of nodes while grouping. If~ancestral relationships are considered, then, for a node, ancestors or descendants will not be grouped together since, by definition, the ancestral history of an ancestor and its descendent is different. Similarly, nodes in the same group will not share any relationship because then their ancestral history will be~different. }
	
	{	To identify nodes with  the same derivation history, first nodes of the same type are grouped, defined as 
		\begin{defi}[{\textbf{Node Grouping}}] \label{def:nodegrouping} \textls[-15]{\emph{Given a provenance graph $G(V,E)$, $\phi=\{G_1,G_2,\ldots,G_k\}$ is a~node-grouping such that}} \\
		\vspace{-12pt}
		\begin{enumerate}[leftmargin=2.3em,labelsep=4mm]
			\item[\emph{(1)}] \emph{$\forall G_i \in \phi, G_i \subseteq {Activity(G)}$ or $G_i \subseteq {Entity(G)}$ , and $G_i \neq \emptyset$,}
			\item[\emph{(2)}] \emph{$\cup_{G_i \in \phi} G_i = V(G)$, }
			\item[\emph{(3)}] \emph{$\forall G_i, G_j \in \phi$ and $(i \neq j), G_i \cap G_j = \emptyset$. }
			\end{enumerate}
		\end{defi}
		In particular, in (1), node grouping is over nodes of two types: activity nodes and entity nodes, in~(2), the union of all node grouping is equal to the nodes in $G$; and, in (3), given a node grouping, groups are not overlapping, but distinct. }

	{We now consider grouping $G$ by ancestry. For this, we identify the ancestors of a node as follows. For a given grouping $\phi$, the ancestors of a node $v$ is the set 
		$Ancestor_{\phi,E}(v) = \{(Ancestor(\phi(u)), Type(u,v)),  (u,v) \in E, Type(u,v) \in \mathcal{L}\}$.  Type of edges is based on types used in W3C PROV standard where $\mathcal{L} = \{used, wasGeneratedBy, wasInformedBy\}$. Nodes that do not have an ancestor are assigned the start node as an ancestor, with a start label edge.  Now, we define grouping nodes by ancestry. 
		\begin{defi}[{\textbf{\HT{A}ncestry grouping}}]\label{def:derivation}
			\emph{A grouping $\phi=\{G_1,G_2,\ldots,G_k\}$ has the same ancestry if it satisfies the following:} 
			\begin{enumerate}[align=parleft,leftmargin=*,labelsep=3mm]
			\item[\emph{(i)}] \emph{Node Grouping Definition \ref{def:nodegrouping}, }
			\item[\emph{(ii)}] \emph{$\forall u,v \in V(G)$, if $\phi(u) = \phi(v)$, then $\forall L_i \in \mathcal{L}$,
			$Ancestor_{\phi,L_i}(u) = Ancestor_{\phi,L_i}(v)$. }
			\end{enumerate}
	\end{defi}}
		
\begin{figure}[H]
		\centering
		\includegraphics[width= 2.5in]{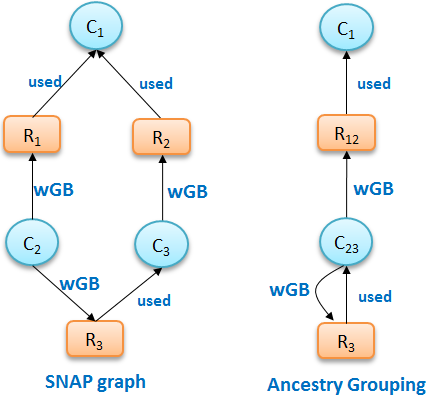}
		\caption{An example of ancestry grouping.}
		\label{fig:SNAPGrouping}
	\end{figure}

	Figure~\ref{fig:SNAPGrouping}\HT{(a)} shows the grouping of nodes due to SNAP, in which nodes with the same types and same ancestors are separated into different groups if their descendants are different, and due to ancestry grouping in Figure~\ref{fig:SNAPGrouping}\HT{(b)}, in which nodes of the same type with same ancestors remain grouped.
	
\begin{figure}[H]
		\centering
		\includegraphics[width=15cm]{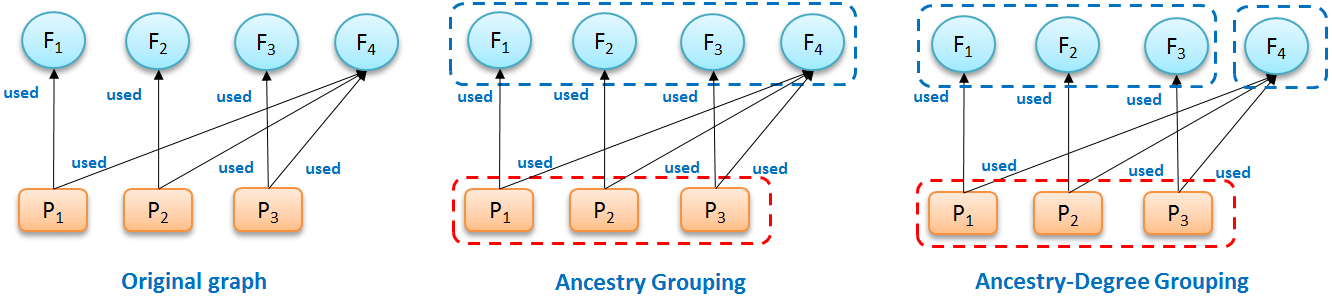}
		\caption{An example of ancestry-degree compatible grouping.}
		\label{fig:ARDCompatible}
	\end{figure}

	{Ancestry grouping, however, may \HT{still} group system files and dependency information together with application-specific nodes. Consider the example in Figure~\ref{fig:ARDCompatible}. Suppose that we have a provenance graph on the left of Figure~\ref{fig:ARDCompatible} 
	that has three nodes of process ($P_1, P_2$ and $P_3$), four nodes of file ($F_1, F_2, F_3$ and $F_4$) and six edges (relationship $used$). The summary ancestry graph shown in center of Figure~\ref{fig:ARDCompatible} dividing the original graph into two groups $g_1 = \{P_1, P_2, P_3\}$ and $g_2=\{F_1, F_2, F_3, F_4\}$, satisfies \ref{def:derivation}. {(All~nodes in every group have similar node types and associate with the same ancestry groups)}. However, from the conceptual point of view, file $F_4$, a dependency used by other processes is different from other file nodes in $g_2$, since this node is the only node that associates with all processes in group $g_1$. In other words, $F_4$ has $InDegree_{g_1,used}(F_4) = 3$ while other nodes in $g_2$ have $InDegree_{g_1,used}(F_1) = InDegree_{g_1,used}(F_2) = InDegree_{g_1,used}(F_3) = 1$.}
	
	{To uniquely differentiate $P_4$, we define ancestry-degree compatible grouping to summarize provenance graphs.
		\begin{defi}[{\bf \HT{A}ncestry-degree grouping}]\label{def:degreederivation}
			\emph{A grouping $\phi=\{G_1,G_2,\ldots,G_k\}$ has the same ancestry-degree if it satisfies the following:} 
			\begin{enumerate}[align=parleft,leftmargin=*,labelsep=3mm]
			\item[\emph{(i)}] \emph{\HT{A}ncestry grouping Definition \ref{def:derivation},} 
			\item[\emph{(ii)}] \emph{$\forall u,v \in V(G)$, if $\phi(u) = \phi(v)$, then $\forall L_i \in \mathcal{L}$, $InDegree_{\phi,L_i}(u) = InDegree_{\phi,L_i}(v)$ and $OutDegree_{\phi,L_i}(u) = OutDegree_{\phi,L_i}(v)$, where $InDegree_{\phi,L_i}(u) = |\{v|v \in \phi, (v, u) \in E, Type(v, u) = L_i\}|$ and $OutDegree_{\phi,L_i}(u) = |\{v|v \in \phi, (u, v) \in E, Type(u, v) = L_i\}|$.}
			\end{enumerate}
	\end{defi}}
	
	\unskip
	
	{Based on this definition, the summary graph shown in Figure~\ref{fig:ARDCompatible} (right) composed of three groups---$g_1 = \{P_1, P_2, P_3\}$, $g_2 = \{F_1, F_2, F_3\}$ and $g_3=\{F_4\}$---is ancestry-degree compatible because it is Ancestry-Grouping compatible and all the nodes in every group have the same number of output edges from/to other groups (all nodes in $g_2$ have one input edge from group $g_1$, all nodes in group $g_1$ have two output edges: one to $g_2$ and one to $g_3$).}\\  \pagebreak
	
	\begin{algorithm}[H]
		\SetKwProg{Def}{}{:}{}
		\SetKwInOut{Input}{Input}
		\SetKwInOut{Output}{Output}
		\Def{\textbf{Ancestry-degree grouping} (
		Labeled G (V, E))}{
			Group all $\{nodes\}$ with same attributes into $\{g_i\}$ and store all groups in a Stack $\Phi=\{g_i\}$\;
			\While {($\Phi$ is not empty)} {
				$g$ = $\Phi$.pop ()\;
				$V_g$ = vertex list of $g$\;
				\textbf{divideGroups} ($\Phi$, $V_g$, ``from'')\;
				\textbf{divideGroups} ($\Phi$, $V_g$, ``to'')\;
			}
			\ForEach {$node$ in $\{nodes\}$}{
				/*groupofNode() returns the group where given node belongs to*/\;
				$\Phi_S$ = $\Phi_S \bigcup$ \textbf{groupofNode} ($node$)\; 
			}
			\Return $\Phi_S$\;
		}
		\Def{\textbf{divideGroups} ($\Phi$, $V_g$, direction)}{
			\ForEach {types of edges or labels}{
				/*$g_i = \{u_i\}$ has at least one vertex $u$ connecting to at least one vertex $v$ in $V_g$*/\;
				\ForEach {group $g_i$ in $\Phi$ and $g_i$ connects to $V_g$}{
					\ForEach {vertex $u$ in group $g_i$} {
						Let $e$ is an edge that connects between vertex $u$ and a vertex $v$ in $V_g$\;
						\If {(direction == "from")}{
							$e$ = $(u, v)$\;
						}
						\Else{
							$e$ = $(v, u)$\;
						}
						Count the number of edges $e$ (i.e., ``degree'' of $u$)\;
					}
					Re-partition all vertices $\{u_i\}$ in group $g_i$ into a list of sub-groups $\{g\_par\}$ with the same degree\;
					\ForEach {group $g\_par_i$  in $\{g\_par\}$}{
						\If{($g\_par_i$ not in $\Phi$)}{
							$\Phi$.push ($g\_par_i$)\;
						}
					}
				}
			}
		}
		\caption{Ancestry-degree grouping} 
		\label{alg:SNAP-i}
	\end{algorithm}
	
	\vspace{12pt}
	{We now describe the summary algorithm, which, given a retrospective provenance graph, produces an ancestry-degree grouping. In this algorithm, we first divide the nodes into group with same node type $\{g_i\}$  and store them in a stack $\Phi$ (Line 2). Next, for each group $g$ in the stack, we re-partition all groups in $\{g_i\}$ in stack by calling function \textbf{divideGroup()} with a list of vertices in $g$ (i.e., Lines 5--7). Function \textbf{divideGroup()} is called twice with different direction parameters, since it is applied on two different directions of edges (i.e., input or output). The main purpose of this function is to re-organize all the groups (i.e., $\Phi_c=\Phi-\{g\}$) that are relevant to $g$ by checking all the vertices and edges of vertices in group $g$ of stack $\Phi$ (Lines 12 and 26). First, for each types of edges or relationships (in~our context, there are three types of edge: $\{used, wasGeneratedBy, wasInformedBy\}$), it calculates the number of edges (or ``degree'') from/to vertices in each group $g_i$ of $\Phi$ to/from a vertex $v$ in $V_g$ (Lines~16--22). Second, it further divides these vertices $u_i$ in each group $g_i$ of $\Phi_c$ by considering the degree of these vertices (i.e., the number of edges from/to vertices in group $g$). This means vertices that belong to the same group must have the same degree (Line 22). Third, we add all the new generated groups $\{g\_par_i\}$ to $\Phi$ if they have never been in $\Phi$ (Lines 24--26), before the next consideration of other groups in $\Phi$. Finally, once all the groups $g$ in $\Phi$ are considered, we obtain the summary graph $\Phi_S$ by joining all the groups together (Lines 8--10).}
	
\begin{figure}[H]
		\centering
		\includegraphics[width=15cm]{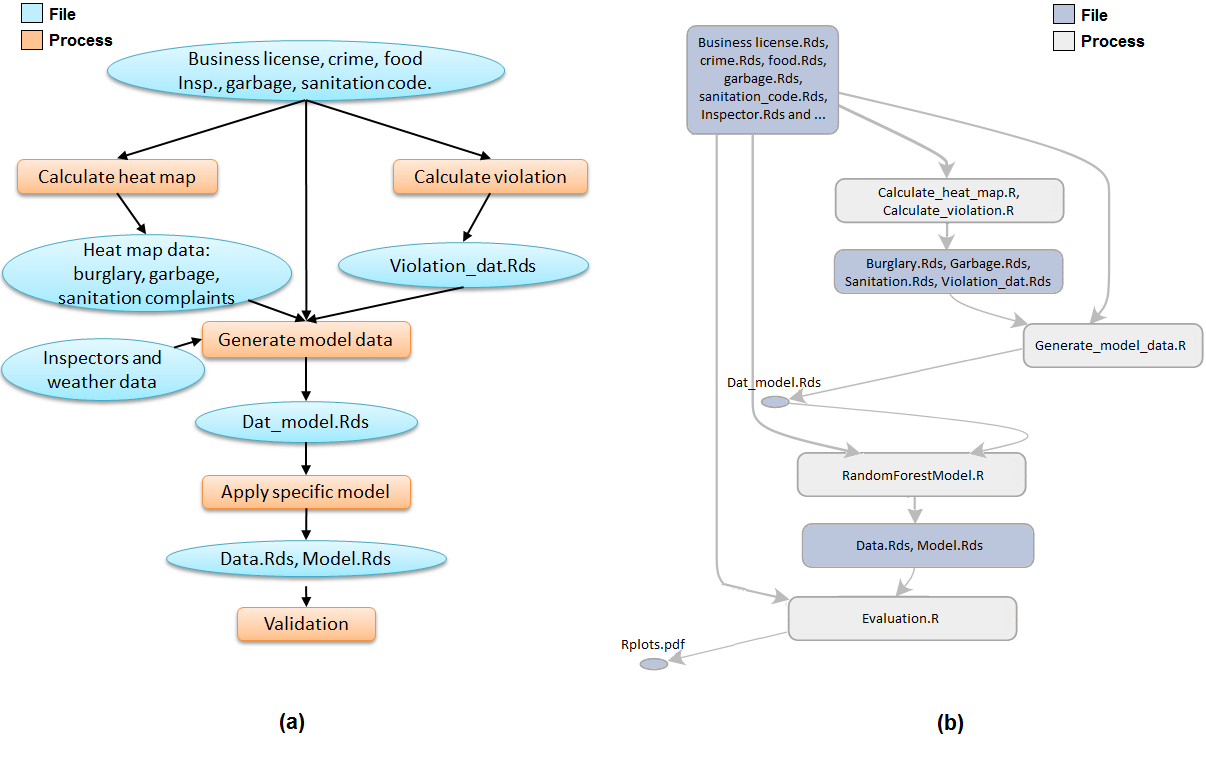}
		\caption{An example of using ancestry-degree grouping. (a) Original user work-flow of FIE and (b) Provenance summarization graph} 
		\label{fig:GraphOptimizationSciunitI}
	\end{figure}

	{Figure~\ref{fig:GraphOptimizationSciunitI} presents an example of graph summarization. The left figure (i.e., Figure~\ref{fig:GraphOptimizationSciunitI}a) shows the work-flow of FIE~\cite{FIE} application drawn by users that describes the conceptual view of FIE application, while the right figure (i.e., Figure~\ref{fig:GraphOptimizationSciunitI}b) shows the provenance summary graph of FIE application after applying ancestry-degree grouping.
	As a general observation, these two graphs are fairly close to each other. The summary graph almost captures all the information about the application that users might need at the general view. There are some minor differences between them. For example, the~two processes ``Calculate heat map'' and ``Calculate violation'' are clearly separate in the left figure while in the right figure they are grouped together. Similarly, the two groups of files ``heat map data'' and ``Violation\_dat.Rds'' are separate in the left figure and grouped in the right. These groups of files and groups of processes are ancestry-degree compatible, and thus they are grouped in the right figure. While they are separate in the application workflow, ancestry-grouping is helpful in general. For~instance, the ancestry-grouping  grouped all data files into a single group at the top.}  

	\section{Experiments}\label{sec:experiments}
	
	The true usefulness of sciunits can only be measured by their adoption. Efficiency of creating sciunits can be a driving force in adopting the use of sciunits over traditional shared research objects. When an efficiently-versioned, easily-created sciunit is shared, along with an embedded, self-describing application workflow, we believe the probability for reuse will greatly increase. In this section, through two complex real-world workflows, we quantify the performance of \HT{containerizing} and repeating sciunits, 
	\HT{and the efficiency} of reusing them utilizing integrated provenance visualizations. We implemented our \HT{\texttt{Sciunit}} client in Python and C. The source code and documentation of \HT{\texttt{Sciunit}} is available from \HT{\url{https://sciunit.run}} \cite{SciunitI:2017:sciunit.run}. 
	
	Sciunit's versioning tool was written in C++, using the block-based deduplication techniques proposed in \cite{Muthitacharoen:ContentBaseddeDuplication} and \cite{Rabin:fingerprinting}. \texttt{sciunit}'s provenance graph visualization was written in Python, using libraries from TensorBoard \cite{Tensorflow}. All~sciunit client \emph{exec} and \emph{repeat} experiments, along with their baseline normal application runs, were conducted on a laptop with an Intel Core i7-4750HQ 2.0 GHz CPU, 16 GB of main memory, and a 1 TB SATA SSD (Solid state disk), running the Arch Linux 
	64-bit OS (at Chicago, Illinois, 60604, USA).
	
	\subsection{Use Cases}\label{subsec:Experimental descriptions}
	
	We consider two real-world use cases for experimental evaluation: (i) the Food Inspection Evaluation ({FIE}) 
	 \cite{FIE} workflow, a computationally-intense use case that has been the running example in our paper, and (ii) \HT{Variable} Infiltration Capacity (VIC) \cite{billah2016using} model, an I/O-intensive (Input/Output-intensive) data pre-processing pipeline for \HT{hydrology} model.

The first use case is notable for its transparency in its rigorous inspection audits, owing to the influence of the Open Data movement within the City of Chicago. The second use case is a highly-relevant test bed for sciunits: the VIC model is very popular in the hydrology community, and its data preprocessing pipeline, which relies heavily on legacy code, is notoriously difficult to reassemble \cite{billah2016using}.
		
\begin{figure}[H]
		\centering
		\includegraphics[width=15cm]{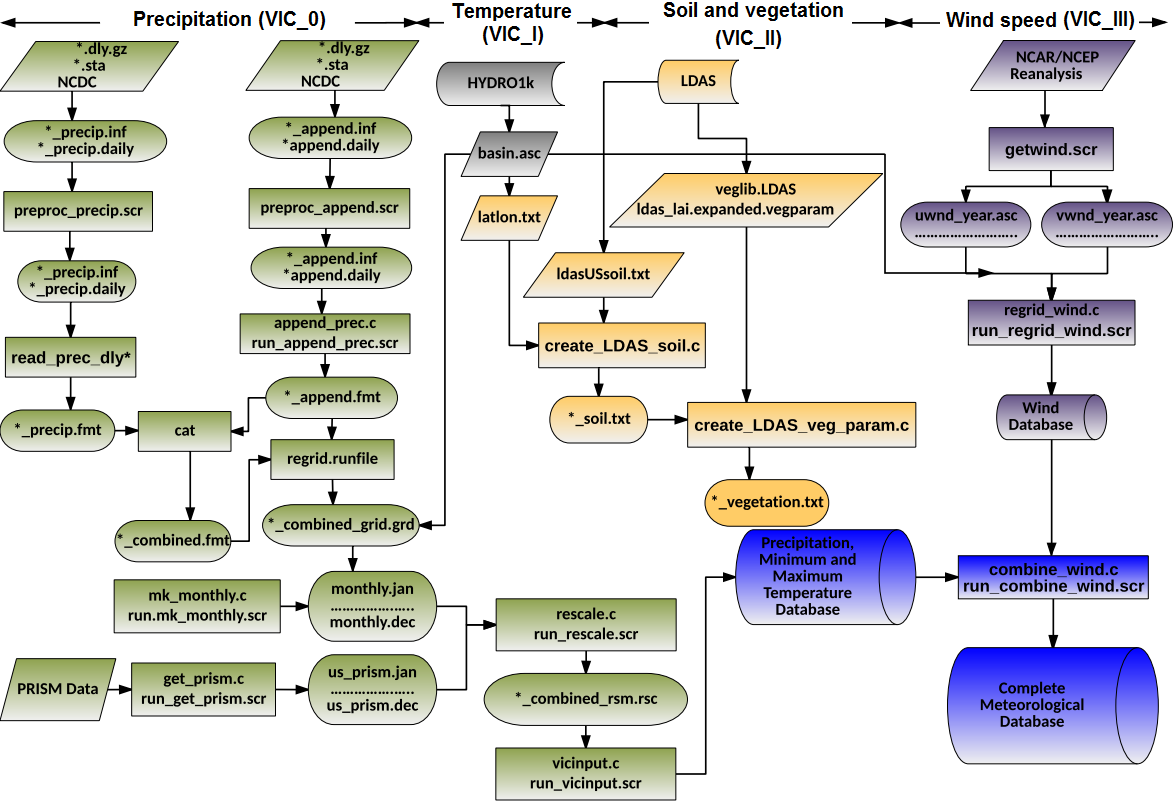}
		\caption{Conceptual view of the VIC workflow~\cite{billah2016using} ((*) in this figure denotes a list of files that share the same prefix or suffix).} 
		
		\label{fig:VIC workflow}
	\end{figure}

	Tables \ref{table:fie application} and \ref{table:vic application} describe the details of FIE and VIC in terms of source code file programming languages, number of source code and data files, number of program files required as dependencies, and total application sizes (both FIE and VIC have four sub-tasks, labeled 0, I, II, and III, which are described below). Figure~\ref{fig:FIE workflow}\HT{(a)} and \HT{Figure}~\ref{fig:VIC workflow} show conceptual views of the application workflows for the two use cases. 
	
	\HT{We assume a sharing} model, in which each step is conducted independently by one user, and subsequently shared with another user who builds upon or forks the shared workflow in the following step. Thus, the FIE workflow, for example, is broken down into the following sub-tasks, each encapsulated in a single application: (i) FIE\_0, which calculates a heat map from downloaded inspection records; (ii) FIE\_I, which processes the heat map to generate data model inputs; (iii) FIE\_II, which applies a specific model and validates it; and (iv) FIE\_III, which downloads the original inspection records and applies an end-to-end validation routine to the previous three sub-tasks. The download process of subtask iv is often the most time-consuming step.

	\begin{table}[H]
		\caption{\HT{Food Inspection Evaluation} sub-task applications.} 
		\label{table:fie application}
		\centering
		\begin{tabular}{ccccc}
			\toprule
			\textbf{} & \textbf{FIE\_0} & \textbf{FIE\_I} & \textbf{FIE\_II} & \textbf{FIE\_III}
			\\ 
			\midrule
			Source code languages 
			& R, Bash & R, Bash & R, Bash & R, Bash
			\\ 
			Source code files & 19 & 20 & 24 & 29
			\\
			Data files & 2 & 8 & 14 & 14
			\\
			Dependency files & 255 & 255 & 411 & 659
			\\ 
			Size of all files & 133.2 MB & 178.4 MB & 289.7 MB & 306.6 MB
			\\
			Normal run time & 52.046 s & 238.833 s & 295.785 s & 7200 s
			\\ 
			\bottomrule
		\end{tabular}
		\vspace{-10pt}
	\end{table}
	\begin{table}[H]
		\caption{\HT{Variable Infiltration Capacity} sub-task applications.} 
		\label{table:vic application}
		\centering
		\begin{tabular}{ccccc}
			\toprule
			\textbf{} & \textbf{VIC\_0} & \textbf{VIC\_I} & \textbf{VIC\_II} & \textbf{VIC\_III}	\\ 
			\midrule
			Source code languages & \multicolumn{4}{c}{C, C++, Python, \HT{C shell script}, Fortran}
			\\ 
			Source code files & 35 & 61 & 77 & 97
			\\ 
			Data files & 3689 & 6313 & 11,460  
			& 11,481
			\\ 
			Dependency files & 247 & 260 & 314 & 357
			\\ 
			Size of all files & 1.2 GB & 1.3 GB & 2.2 GB & 2.3 GB
			\\ 
			Normal run time & 158.734 s & 306.069 s & 363.147 s & 377.29 s
			\\ 
			\bottomrule
		\end{tabular}
		\vspace{-10pt}
	\end{table}
	
	\subsection{Creating Sciunits}
	\label{subsec:Reproducing overhead}
	
	Tables \ref{table:fie application} and \ref{table:vic application} present the baseline normal execution times for the sub-tasks of the two use cases. We note that each application encompasses substantial resources (in the form of code and data), has many external dependencies, and is also characterized by lengthy CPU-and-memory-intensive tasks. Additionally, the nature of FIE's processing tasks differ significantly from those of VIC. FIE~front-loads its input data sets into memory, and then utilizes machine-learning logic to process its data. VIC~also runs many intricate calculations, but differs from FIE in that it interlaces file input and output operations regularly throughout its code. This difference \HT{is} key in understanding that sciunits have minimal performance impact on most---but not all---types of applications.

\begin{figure}[H]
		\centering
		\includegraphics[width=3.75in]{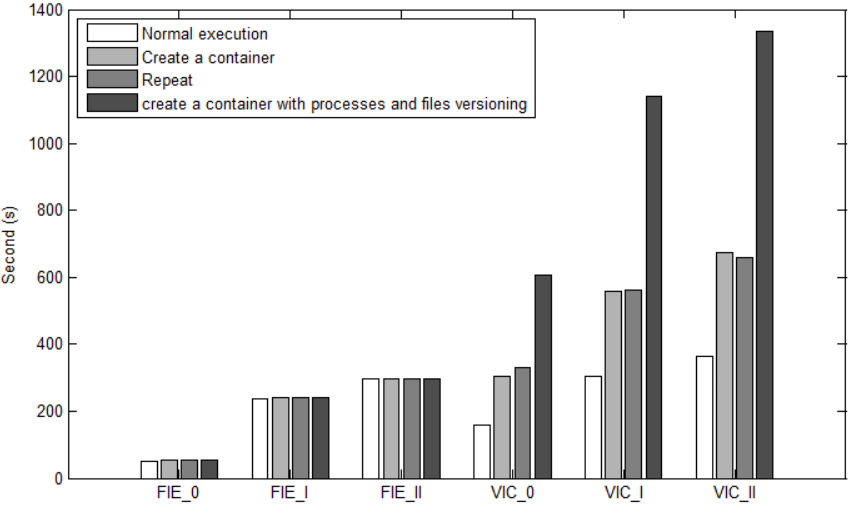}
		\caption{Execution times for normal runs, creating containers, and repeating.}
		\label{fig:overhead}
	\end{figure}

	Figure \ref{fig:overhead} compares the baseline normal execution time of each subtask with the time consumed by packaging the sub-task with the \texttt{sciunit}'s \emph{exec} command, and with the time consumed by repeating the sub-task with the \texttt{sciunit}'s \emph{repeat} command. \HT{Test results for the FIE\_III and VIC\_III sub-tasks were omitted due to significant amounts of network-dependent downloading operations.} We note that the performance impact of auditing and repeating on FIE's run times was negligible: auditing FIE with \emph{exec} resulted in only a 3.6\% time increase, and executing FIE with \emph{repeat} added only a 1.3\% increase to run time. 
	\HT{Meanwhile}, in FIE, the I/O access time is much less than CPU processing time. Also note that our tests were done on SSD offering much better performance than HDD (hard disk drive). 
The reasons explain why the overheads in these cases are negligible. Conversely, \HT{containerizing and repeating} VIC with \HT{\texttt{Sciunit}} nearly doubled the original application run times: as noted in the preceding paragraph, it was evident that using \HT{\texttt{Sciunit}} with I/O-intensive (Input/Output-intensive) applications affected application performance \HT{significantly}. 
			
	We obtain one further observation from these experiments by comparing each application \emph{exec} time with its corresponding \emph{repeat} time. Compared to application repeat increases, auditing increases were slightly higher. This difference can be understood by examining sciunit's behavior during AV audit-time: auditing entails copying an application's code and data into a sciunit container, but running the sciunit container with \emph{repeat}, however, only redirects to these copied files, and therefore precludes the file copy time.	
	
	{\subsection{Repeatability Evaluation}\label{subsec:repeatabilityEvaluation}
		To measure the exact repeat execution, we run all test cases presented in Tables~\ref{table:fie application} and \ref{table:vic application} on many different environments such as \HT{Ubuntu, Arc Linux, CentOS 7, Fedora 26, RHEL 7 or Debian.}
		 The results are shown in Table~\ref{table:reapeatability evaluation}.}
	
	{As clearly shown in Table~\ref{table:reapeatability evaluation}, these applications can be repeated successfully in all tested environments. We also applied Algorithm~\ref{algorithm:similar graph} to verify the provenance graph isomorphism between original runs and re-executions. The results are recorded in the column ``Provenance graph isomorphism''. Since our isomorphism algorithm will finish when the first bijection is found, its performance is good even for large provenance graphs (e.g., it takes less than one second for handling with provenance graphs having up to 150 nodes and 320 edges).}

	\begin{table}[H]
		\caption{Exact repeatability evaluation.}
		\label{table:reapeatability evaluation}
		\centering
	\scalebox{0.8}[0.8]{
	\begin{tabular}{m{2cm}<{\centering}m{5cm}<{\centering}m{3cm}<{\centering}m{2cm}<{\centering}m{3cm}<{\centering}}
			\toprule
&  \multicolumn{2}{c}{\multirow{2}{*}{\textbf{Repeatability}}}	& 	\multicolumn{2}{c}{\textbf{Execution Time (s)  (Measured on }}\\
& &	& 	\multicolumn{2}{c}{\textbf{Our above Described Machine)}}\\
 \midrule
\textbf{Selected	  Applications} & \textbf{Re-execution (Tested on \HT{Ubuntu, Arc Linux, CentOS 7, Fedora 26, RHEL 7 or Debian})} & \textbf{Provenance Graph Isomorphism} &\textbf{ Execution} & \textbf{Repeat}\\ 
\midrule		
			FIE\_0 
			 & Succeed & Matched & 52.046 & 53.954
			\\ 
			FIE\_I & Succeed & Matched & 238.833 & 240.014
			\\ 
			FIE\_II & Succeed & Matched & 295.785 & 297.681
			\\ 
			FIE\_III & Succeed & Matched & \HT{7,200} & \HT{7,308.6}
			\\ 
			VIC\_0 & Succeed & Matched & 158.734 & 606.21
			\\ 
			VIC\_I & Succeed & Matched & 306.069 & \HT{1,140.5}
			\\ 
			VIC\_II & Succeed & Matched & 363.147 & \HT{1,332.98}
			\\ 
			VIC\_III & Succeed & Matched & 377.29 & \HT{1,384.89}
			\\
		\bottomrule
		\end{tabular}}
	\end{table}

	{\subsection{Partial and Modified Repeat Execution}\label{subsec:partial and modified repeat}
		The main ideas of partial reproducibility are to reduce both execution time (only execute the necessary parts) and container size (not to include the data files or dependencies that will not be used). Therefore, we measure the partial reproducibility by the following criteria: (i) correctness; (ii) resource usability, and (iii) execution time. }
	
	{Table~\ref{table:partial and modified repeat executions} shows our evaluation on partial and modified reproducibility on the two selected use cases (i.e., FIE\_III and VIC\_III). In particular, we built the partial and modified containers on the originals of FIE\_III and VIC\_III. In FIE\_III, we selected only the process ID that calculates the heat map from the downloaded file, and then built the sub-container (i.e., FIE\_Par) using Algorithm~\ref{algorithm:build sub-container}. \HT{In particular, we note that in this experiment, Algorithm~\ref{algorithm:build sub-container} used only direct descendants to build partial containers. A more general experiment using all descendants is in accompanying technical report~\cite{TechnicalReport:2018}}. Meanwhile, the modified execution of FIE\_III (i.e., FIE\_Mod) was tested when we change the inputs (i.e., use new weather data file: ``/tmp/weather\_201801.Rds'') using the \emph{given} command (see Section~\ref{sec:architecture}). Similarly, in~VIC\_III, we built the partial container (VIC\_Par) with the process ID that only processes precipitation data. Table~\ref{table:partial and modified repeat executions} shows the number of files and dependencies within the partial containers in comparison with those of original containers, as well as the differences in runtimes of repeat and original run. Values from row \HT{``\# of files not used"} denote that all files in the partial containers are touched when the application runs, indicating no \HT{extra file was included using Algorithm~\ref{algorithm:build sub-container} in these partial containers. Meanwhile, row ``Executable'" shows if partial and modified repeatability was successful.}
	
	\begin{table}[H]
		\caption{\HT{Partial and Modified Repeat Executions.}}
		\label{table:partial and modified repeat executions}
		\centering
			\scalebox{0.8}[0.8]{
		\begin{tabular}{m{2.8cm}<{\centering}m{2.2cm}<{\centering}m{2.2cm}<{\centering}m{2.2cm}<{\centering}m{2.2cm}<{\centering}m{2.2cm}<{\centering}m{2.2cm}<{\centering}}
			\toprule
&\multicolumn{3}{c}{\textbf{Original, Modified and Partial   Executions of FIE\_III}} &		\multicolumn{3}{c}{\textbf{Original, Modified and Partial Executions of VIC\_III}}		
			\\ \midrule
			\textbf{Information} 
			& \textbf{FIE\_III} & \textbf{FIE\_Mod} & \textbf{FIE\_Par} & \textbf{VIC\_III} & \textbf{VIC\_Mod} & \textbf{VIC\_Par}
			\\ 
			\midrule	
			\# of executable files  
			& 29 & 29 & 19 & 97 & 97 & 35
			\\ 
			\# of data files & 14 & 14 & 2 & 11,481  
			& 11,481 & 36
			\\ 
			\# of dependencies & 659 & 659 & 255 & 357 & 357 & 247
			\\ 
			Executable & Succeed & Succeed & Succeed & Succeed & Succeed & Succeed
			\\ 
			Execution time (s) & \HT{7,200} & - 
			 & 52 & 377 & - & 159
			\\ 
			\# of files not used & 0 & - & 0 & 0 & - & 0
			\\ 
			\bottomrule
		\end{tabular}}
	\end{table}

	\subsection{Reusing Sciunits with Provenance Visualizations}\label{subsec:Graphic performance}
	
	Application virtualization has traditionally led to fine-grained provenance graphs that are often difficult to decipher. In this sub-section, we determine if our summarization rules produce a usable provenance graph that is closer to a theoretical, intuitive user application workflow. We focus this discussion on experiments for the FIE sub-tasks, but \HT{mention} that experiment results for the VIC sub-tasks were similar.
	
	To evaluate the effectiveness of summarization, we first considered three traditional, replete (i.e.,  fine-grained) provenance \HT{traces} generated by \HT{\texttt{Sciunit}} on
auditing FIE\_I, FIE\_II, FIE\_III {(We did not consider FIE\_0 in this analysis since its original replete graph was too small and simple to benefit measurably from summarization)}. We calculated the number of nodes (each a process or a file) and edges present in each replete graph. Next, we calculated the number of nodes present in the corresponding \HT{sciunit} container provenance graphs. \HT{These graphs were summarized by using both the similarity and packability rules (i.e., collapsing retrospective provenance method) and ancestry-degree grouping method}. Figure~\ref{fig:Number Objects} depicts a comparison of these methods (i.e., original, collapsing retrospective provenance method and ancestry-degree grouping method). 
	
	\begin{figure}[!]
		\centering
		\includegraphics[width=3.5in]{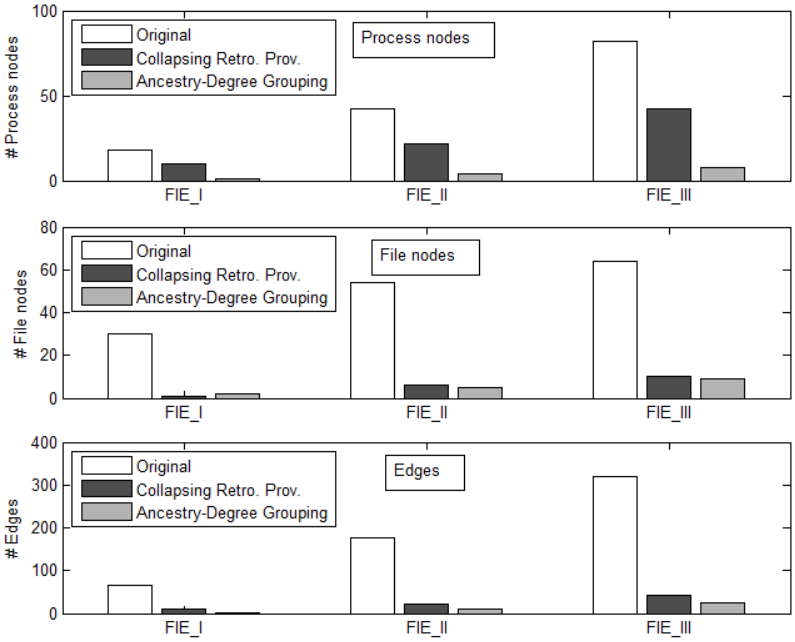}
		\caption{Number of nodes and edges in original and summarized graphs.}
		\label{fig:Number Objects}
	\end{figure}
	
	Graph summarization reduced the number of file nodes, process nodes, and edges by averages of 88\%, 41\%, and 87\% with graphs generated by collapsing retrospective provenance method and 90\%, 91\% and 93\% with graphs generated \HT{by ancestry-degree grouping method.} 
	
	We also measured the number of clicks needed to expand summarized graphs to replete graphs. For FIE\_III, which had the largest graph, expanding \textit{any} summarized node required a maximum of four user clicks to reach its replete view. Expanding \textit{all} the nodes in this large graph took 45 clicks. This observation showed that graphs were  summarized very well spatially and intuitively, yet still capable of allowing fully-detailed provenance examination with a modest amount of user interaction.
	
	As seen in Figure~\ref{fig:Number Objects}, there are some differences between summary graphs from the collapsing retrospective provenance method and ancestry-degree grouping method. In general, applying the ancestry-degree grouping method is more efficient than collapsing retrospective provenance method in terms of number of objects. However, the graphs from collapsing retrospective provenance method are still clear and it is easy to understand the system detailed information. Indeed, the key differences between these two are the messages they deliver (see summary graphs in Figures~ \ref{fig:Graph Optimization} and \ref{fig:GraphOptimizationSciunitI}). The summary graph from collapsing retrospective provenance method describes how an application be executed with its dependencies. Meanwhile, the one from ancestry-degree grouping method illustrates the conceptual view of an application. Therefore, we extend the new summarization method while keeping the old version and let users select between these methods according to which information they would prefer to examine.
	
	\section{Conclusions}\label{sec:conclusion}
	
	Computational reproducibility \cite{Freire:2012} is a formidable goal requiring advancements in policy \cite{Victoria:2013:Toward}, user perception \cite{Penny:2016:Nature}, and reproducible practices and tools \cite{Stodden2014}. As we embrace this goal within the sciences \cite{Malik:2017:geotrusthub}, we have encountered that computational provenance is the key to enhancing the experience of reproducible packages as created by the use of application virtualization. In this paper, we have outlined methods to create and store containers based on application virtualization and demonstrated an easy-to-use Git-like client, the \HT{\texttt{Sciunit}} that enables reproducibility for a wide variety of use cases. We showed how embedded provenance can be used to reuse the sciunit and understand them by summarizing embedded provenance. The field of computational reproducibility is a moving target and there are {emerging requirements to use provenance to address reproducibility within Jupyter notebooks~\cite{Ma:2017:Provenance} , Matlab, distributed data-intensive programs, and parallel HPC applications}, which we hope to address as part of future work.

	
	\vspace{6pt}
	
	\acknowledgments{The authors would like to thank Tom Schenk and Gene Leynes for the FIE use case and Jonathan Goodall and Bakinam Essawy for the VIC use case. The authors would also like to acknowledge support for this work from the National Science Foundation under grants NSF ICER-1639759, ICER-1661918, ICER-1440327, and ICER-1343816. 
	} 

\authorcontributions{All authors contributed equally to the paper. Particularly, Z.Y., G.F., and T.M. designed and implemented \texttt{Sciunit} and provenance auditing. D.H.T.T. and T.M. developed  algorithms for exact, partial, and modified repeat executions. D.H.T.T and G.F performed the experiments and analyzed the data. S.K. designed the prospective provenance summarization technique. T.M and D.H.T.T primarily wrote the paper.}	 

\conflictsofinterest{The authors declare no conflict of interest.}  
	
	
	
	\reftitle{References}

\end{document}